\documentclass{pasj00}
\begin{document}
\SetRunningHead{N. Isobe, et al. }
               {Suzaku investigation into the nature of M33 X-8}
\Received{2012/02/27}
\Accepted{2012/05/20}

\title{
Suzaku investigation into the nature 
of the nearest ultraluminous X-ray source, M33 X-8. 
}

\author{%
Naoki       \textsc{Isobe}      \altaffilmark{1},
Aya         \textsc{Kubota}     \altaffilmark{2},
Hiroshi     \textsc{Sato}       \altaffilmark{2},
and 
Tsunefumi   \textsc{Mizuno}     \altaffilmark{3}
}
\altaffiltext{1}{Institute of Space and Astronautical Science (ISAS), 
        Japan Aerospace Exploration Agency (JAXA) \\ 
        3-1-1 Yoshinodai, Chuo-ku, Sagamihara, Kanagawa 252-5210, Japan}
\email{n-isobe@ir.isas.jaxa.jp}
\altaffiltext{2}{Department of Electronic Information Systems,
   Shibaura Institute of Technology, \\
   307 Fukasakum Minuma-ku, Saitama-shi, Saitama, 337-8570, Japan}
\altaffiltext{3}{Department of Physics, Hiroshima University, 
   1-3-1 Kagamiyama, Higashi-Hiroshima, Hiroshima 739-8526, Japan}
\maketitle
\begin{abstract}
The X-ray spectrum of the nearest ultraluminous X-ray source, M33 X-8, 
obtained by Suzaku during 2010 January $11$ -- $13$, 
was closely analyzed to examine its nature. 
It is, by far, the only data with the highest signal statistic 
in $0.4$ -- $10$ keV range.
Despite being able to reproduce the X-ray spectrum, 
Comptonization of the disk photons failed 
to give a physically meaningful solution.
A modified version of the multi-color disk model, 
in which the dependence of the disk temperature on the radius 
is described as $r^{-p}$ with $p$ being a free parameter, 
can also approximate the spectrum.
From this model, 
the innermost disk temperature and bolometric luminosity were obtained as 
$T_{\rm in} = 2.00_{-0.05}^{+0.06}$ keV and 
$L_{\rm disk} = 1.36 \times 10^{39} (\cos i)^{-1}$ ergs s$^{-1}$,
respectively, where $i$ is the disk inclination.
A small temperature gradient of $p = 0.535_{-0.005}^{+0.004}$,
together with the high disk temperature, 
is regarded as the signatures of the slim accretion disk model, 
suggesting that M33 X-8 was accreting at high mass accretion rate.
With a correction factor for the slim disk taken into account,
the innermost disk radius, $R_{\rm in } =81.9_{-6.5}^{+5.9} (\cos i)^{-0.5}$ km,
corresponds to the black hole mass of $M \sim 10 M_\odot (\cos i)^{-0.5}$.
Accordingly, the bolometric disk luminosity is estimated to be about
$80 (\cos i)^{-0.5}$\% of the Eddington limit.
A numerically calculated slim disk spectrum
was found to reach a similar result.
Thus, the extremely super-Eddington luminosity is not required 
to explain the nature of M33 X-8.
This conclusion is utilized to argue for the existence of
intermediate mass black holes with $M \gtrsim 100 M_\odot$
radiating at the sub/trans-Eddington luminosity,  
among ultraluminous X-ray sources 
with $L_{\rm disk} \gtrsim 10^{40}$ ergs s$^{-1}$. 
\end{abstract}

\section{Introduction} 
\label{sec:intro}
A number of nearby galaxies are known to host luminous X-ray sources 
with an X-ray luminosity of $L_{\rm X} \gg 10^{39}$ ergs s$^{-1}$
(e.g., \cite{ULX_Einstein,ULX_complete_sample}).
They are usually referred to as ultraluminous X-ray sources 
(ULXs; \cite{ULX_asca}).
The ULXs are found to be less common in the Local Universe,
as no persistent ULX is yet identified in our Galaxy and M31.  
In spite of significant efforts in more than three decade 
since their discovery,
we have not yet arrived at the general conclusion on their nature. 

It is widely believed from their observational features (e.g., \cite{ULX_asca})
that the ULX phenomena are related to accreting black holes. 
Because the luminosity of an accreting black hole is though to increase 
as a function of the black hole mass $M$ and accretion rate $\dot{M}$ 
on the simplest assumption, 
two distinct ideas are currently proposed 
to interpret the high X-ray luminosity of the ULXs. 
The first option supposes that the luminosity of the ULXs 
should not exceed the Eddington luminosity ($L_{\rm E}$), 
as most Galactic black hole binaries meet this criterion 
(e.g., \cite{XTEJ1550}). 
As a result, the ULXs are required to harbor 
an intermediate-mass black holes with a mass of $M \gg 10 M_\odot$
with $M_\odot$ being the solar mass. 
In the second alternative, a super Eddington luminosity is permitted.
Consequently, the ULXs are regarded as a stellar-mass black hole 
($M \lesssim 10 M_\odot$), 
accreting at a super-critical mass accretion rate 
($\dot{M} \gg L_{\rm E}/c^2$ with $c$ being the speed of light). 

It is of vital importance to construct a reliable technique
to gauge the mass and/or mass accretion rate of the ULXs,
in order to distinguish these two conflicting interpretations.
For the purpose, the dynamical mass determination 
is thought to provide inevitably the most stringent constraint,
as is the case of the Galactic black hole binaries 
(e.g., \cite{GROJ1655_mass,GX339_mass}).
However, the dynamical mass of the ULXs has remained yet to be explored.

X-ray spectra and their variability have been 
one of the most valuable tools to probe the properties of the ULXs. 
A naive inspection on the X-ray spectra in the $0.5$ -- $10$ keV range 
categorizes the ULXs into two types;
those with an X-ray spectrum dominated by emission from accretion disk 
which is approximated by the multi-color disk (MCD; \cite{MCD1,MCD2}) model,
and those exhibiting a power-law (PL) like X-ray spectrum. 
In addition, a growing number of ULXs are reported 
to make a spectral transition between the MCD-like and PL-like states 
(e.g., \cite{IC342_transition,NGC2403Src3}).
These spectral behaviors apparently resemble 
those of Galactic black hole binaries 
in the classical high/soft and low/hard states. 
However, several observational obstacles are known for this assumption
as follows (e.g., \cite{ULX_asca,ULX_asca2}).
The X-ray spectrum of the MCD-like ULXs requires 
a flatter radial temperature distribution within the disk (e.g., \cite{M81X9}) 
in comparison with that in the standard accretion disk \citep{standard_disk}
which is adopted in the MCD model. 
The innermost disk radius $R_{\rm in}$ of the MCD-like ULX 
appears to be variable as $R_{\rm in} \propto T_{\rm in}^{-1}$ 
with $T_{\rm in}$ being the innermost disk temperature 
(e.g., \cite{ULX_asca2,NGC2403Src3}),
while the Galactic black hole binaries in the high/soft state 
are reported to follow a constancy of $R_{\rm in}$ 
(e.g., \cite{LMCX-3_Constant_Rin}) 
which indicates that $R_{\rm in}$ corresponds 
to the last stable circular orbit of the accretion disk.
The photon index of the PL-like ULXs 
($\Gamma > 2$; e.g., \cite{IC342_VHS,NGC2403Src3}) 
tends to be larger than those of the Galactic black hole binaries 
in the low/hard state ($\Gamma \sim 1.5$; \cite{Index_in_low/hard}). 

Recent progresses in observational (e.g., \cite{XTEJ1550,XTEJ1550_VHS}) 
and theoretical (e.g., \cite{slimdisk}) studies 
on Galactic black hole binaries and their accretion disk,
especially near the highest end of their X-ray luminosity,  
newly identified two spectral states called the very high and slim disk ones,  
in addition to the classical low/hard and high/soft states. 
In the very high state, 
the Galactic black hole binaries are reported to exhibit 
a steep PL spectrum with a photon index of $\Gamma \gtrsim 2.4$, 
which is softer than the PL component observed  
in the high/soft state ($\Gamma = 2$ --$2.2$).
The steep PL spectrum in this situation 
is ascribed to Comptonization of the disk photons within the hot corona
surrounding the accretion disk.
In contrast, 
when the Galactic black hole binaries become brighter than 
in the very high state,
they are reported to enter the slim disk state
in which a disk dominant X-ray spectrum 
with a flat disk temperature gradient
was observed,
in addition to an apparent anti-correlation between  
the innermost disk radius and temperature \citep{XTEJ1550}. 
These properties are successfully explained 
by the theory of a slim accretion disk,
where an optically-thick advection and photon trapping 
\citep{photon-trapping} effectively work 
due to a high accretion rate ($\dot{M} > L_{\rm E} / c^2$).

Triggered by these findings,
it is proposed that the MCD-like and PL-like ULXs reside 
in the slim disk and very high states, respectively
(e.g., \cite{ULX_asca2,IC342_VHS,M81X9,NGC1313_Suzaku,NGC2403Src3}).
If this idea is systematically proved to be convincing, 
we will have acquired an empirical method 
to restrict the black hole mass of the ULXs.
This is because the X-ray luminosity of the Galactic black hole binaries 
in these states are reported typically 
as $L_{\rm X} = (0.3$ -- $1) L_{\rm E}$ 
(e.g., \cite{XTEJ1550,GROJ1655_VHS}).
In other word, this interpretation consequently indicates 
that the ULXs are shining at a sub- or trans-Eddington luminosity.
Independently, based on numerical simulation,
\citet{Mass_Slimdisk} extended to the slim accretion disk
the mass estimation from the innermost disk radius, 
which is widely accepted for Galactic black hole binaries 
in the high/soft state accompanied with the standard accretion disk. 
It is important to confirm 
observationally the consistency between these techniques.

Recently, it is verified that Suzaku \citep{Suzaku} is very useful 
to investigate the X-ray spectral properties of the ULXs
\citep{NGC1313_Suzaku,M82X1_Suzaku,SuzakuJ1305,NGC2403Src3}. 
These results motivated us to perform a Suzaku observation 
of the nearest ULX, M33 X-8 \citep{M33_src},  
in which its X-ray spectrum was obtained with the highest signal statistics
in the $0.4$ -- $10$ keV range. 
In the present paper,
we utilized extensively the the Suzaku data of the source 
to investigate the nature of the ULXs. 
 
\begin{figure}[t]
\begin{center}
\FigureFile(80mm,80mm){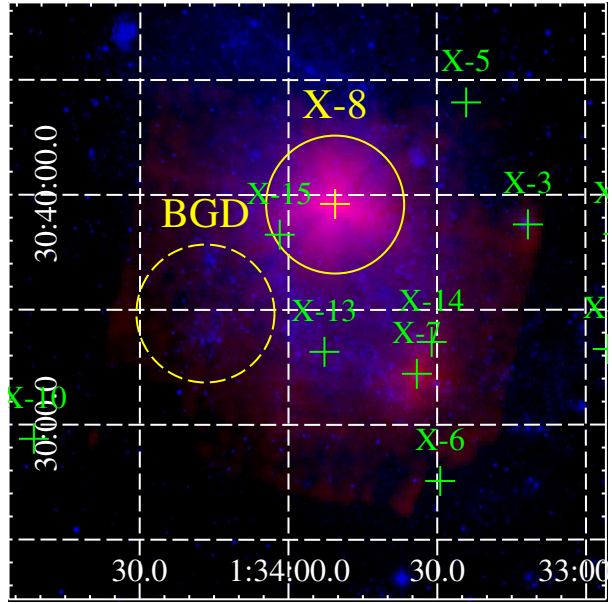}
\end{center}
\caption{
Background-inclusive Suzaku XIS image in the range of 0.5 -- 10 keV 
of M33 (red) with a bin size of $10" \times 10"$, 
on which the optical image taken with the Digitized Sky Survey (blue) 
is superposed.
The data from all the active XIS chips (XIS 0, 1 and 3; \cite{XIS})
were summed up. 
The XIS image was smoothed with a two-dimensional Gaussian kernel 
of a $20"$ radius,
while it is not corrected for the exposure. 
The regions irradiated by the radioactive calibration sources
at the corners of the CCD chips were removed. 
The X-ray sources detected with ROSAT 
are indicated with the green crosses, 
accompanied by the source name \citep{M33_src}. 
The XIS signals from M33 X-8 and background are integrated 
within the solid and dashed circles, respectively. 
}
\label{fig:image}
\end{figure}

\section{Observation and Data Reduction} 
\label{sec:obs}
Within $\sim 20$ arcmin from the center of the nearby normal galaxy, M33, 
more than 10 bright X-ray sources were discovered 
with the Einstein observatory \citep{M33_src}. 
Among these sources, X-8 located near the center of the galaxy 
is known to show the ULX characteristics \citep{ULX_asca}. 
Thanks to it proximity at a distance of 795 kpc \citep{M33_distance},
it extensively studied with the previous and active instruments
including ASCA \citep{M33X8_ASCA}, BeppoSAX \citep{M33X8_SAX}, 
Chandra \citep{M33X8_Chandra}, and XMM-Newton (e.g., \cite{M33X8_XMM}).
The typical X-ray luminosity of M33 X-8, 
$L_{\rm X} \gtrsim 10^{39}$ ergs s$^{-1}$, is at the lowest end of those of ULXs;
only the Galactic microquasar, GRS 1915+105, 
is confirmed to reach occasionally 
such a high X-ray luminosity value (e.g., \cite{GRS1915_Suzaku}).
Some authors claimed a flat disk temperature distribution 
to suggest the slim disk interpretation for the source 
\citep{M33X8_slimdisk,M33X8_XMM},
although several conflicting ideas were proposed 
\citep{M33X8_XMM2,M33X8_LowTeCompton}.
These make M33 X-8 one of the most important object 
to disentangle the nature of the ULXs 
by connecting the properties of the ULXs 
with those of the Galactic black hole binaries. 

The Suzaku observation of M33 X-8 was performed on 2010 January 11 -- 13.
The X-ray imaging spectrometer (XIS; \cite{XIS}) and 
hard X-ray detector (HXD; \cite{HXD}) was operated 
in the normal clocking mode with no window option 
and in the normal mode, respectively. 
The source was placed at the HXD nominal position 
of the X-ray telescope \citep{XRT} above the XIS. 
Unfortunately, 
due to a significant increase in the thermal noise of the HXD PIN,
the HXD effective exposure was found to reduce to only about 6.1 ks. 
Therefore, we concentrate on the XIS data in the present paper. 

The data reduction was performed 
with the standard software package HEADAS 6.10. 
We reprocessed the XIS data with the HEADAS tool {\tt aepipeline},
by referring to the calibration database (CALDB) 
as of 2011 February 10.
In accordance with the standard manner, 
the following filtering criteria were adopted; 
the spacecraft is outside the south Atlantic anomaly (SAA),
the time after an exit from the SAA is larger than 436 s,
the geometric cut-off rigidity is higher than $6$ GV, 
the source elevation above the rim of bright and night Earth is 
higher than $20^\circ$ and $5^\circ$, respectively, 
and the XIS data are free from telemetry saturation.
As a result, we derived 91.0 ks of good exposure.
The XIS events with a grade of 0, 2, 3, 4, or 6 
were picked up for the scientific analysis below.

\section{Results} 
\label{sec:results}
\subsection{X-ray image} 
\label{sec:image}
Figure \ref{fig:image} displays the 0.5 -- 10 keV XIS image of M33 X-8,
which is superposed on the optical image of the host galaxy
taken from the digitized sky survey \citep{dss}.
The brightest X-ray source in the image, 
detected at the center of the host galaxy, corresponds M33 X-8.
A systematic offset of $15.6''$ was noticed 
in the XIS coordinate determination,
while it is found to be within the current systematic uncertainties 
\citep{Suzaku_pos_acc}. 
We thus re-calibrated the XIS coordinate by ourselves,
referring to the ROSAT position of M33 X-8 itself  
(($\alpha_{\rm 2000}$, $\delta_{\rm 2000}$) 
= (\timeform{01h33m50.58s}, \timeform{+30d39m36.7s}); \cite{M33_src}). 
Within the XIS field of view, another X-ray source M33 X-7,
a candidate of a massive spinning black hole 
with a mass of $M = (15.6 \pm 3.2) M_\odot$ \citep{M33X7}, 
is clearly seen at a position consistent with the ROSAT one 
(($\alpha_{\rm 2000}$, $\delta_{\rm 2000}$) 
= (\timeform{01h33m34.08s}, \timeform{+30d32m13.2s}); \cite{M33_src}). 
However, due to the insufficient signal statistics from the XIS, 
we regard its X-ray spectral properties 
as irrelevant to the scope of the present paper.

In the scientific analysis below, 
the XIS signals from M33 X-8 were accumulated 
within a solid circle in figure \ref{fig:image} 
with a $3'$ radius centered on the source,
while the background level was evaluated
from a source free region with the same radius, 
denoted by the dashed circle in figure \ref{fig:image}.
Although there is a faint ROSAT X-ray source, M33 X-15, 
near the edge of the source integration region, 
we estimated its contamination as negligible
in comparison with the X-ray flux from M33 X-8 (less than $1$\%),
based on the ROSAT result \citep{M33_src}.

\begin{figure}[t]
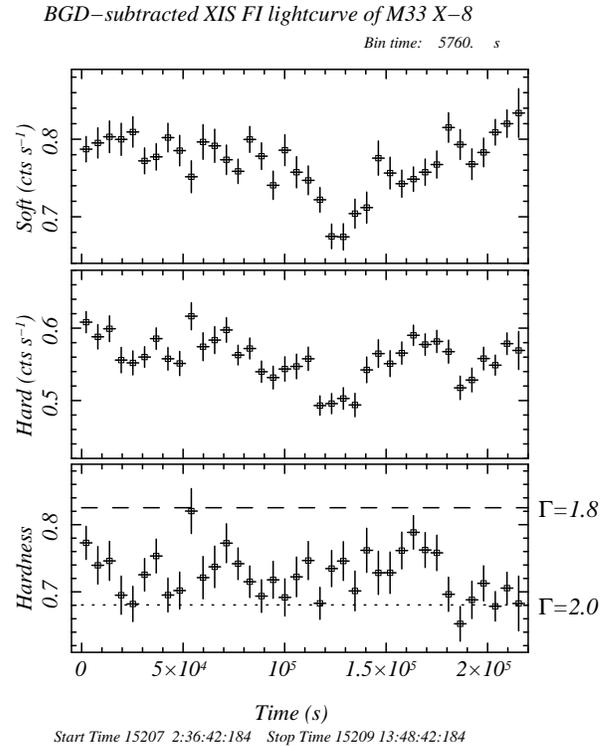

\begin{center}
\FigureFile(80mm,80mm){figure2.ps}
\end{center}
\caption{Background-subtracted XIS FI lightcurve of M33 X-8.
Each time bin is set to 5760 s,
corresponding to the orbital period of Suzaku. 
The top and middle panels show 
the count rate in the soft (0.5 -- 2 keV) and hard bands,
respectively.  
The source hardness, 
simply evaluated as the hard-to-soft count-rate ratio,
is shown in the bottom panel. 
Assuming the Galactic absorption toward M33, 
$N_{\rm H} = 1.1 \times 10^{21}$ cm$^{-2}$ \citep{NH},
a simple PL model with a photon index of $\Gamma = 2.0$ and $1.8$
corresponds to the hardness shown with the dotted and dashed lines, 
respectively. 
}
\label{fig:lcurve}
\end{figure}

\begin{table*}[t]
\caption{Summary of the model fitting to the XIS spectra of M33 X-8}
\label{table:spec}
\begin{center}
\begin{tabular}{llllll}
\hline\hline 
Model                 & MCD\footnotemark[$\|$] & PL\footnotemark[$\|$] 
                      & MCD+PL                &  MCD+THCOMP        & extended MCD  \\ 
\hline 
$F_{\rm X}$\footnotemark[$*$]
                      & $ 1.50 $              & $1.60$                
                      & $1.57$                & $ 1.55 $           &  $1.56$        \\
$L_{\rm X}$\footnotemark[$\dagger$]
                      & $0.57$                & $0.83$                
                      & $0.75$                & $0.65$              & $0.71$        \\
$N_{\rm H}$ ($10^{21}$ cm$^{-2}$)  
                      & $ 0.0 $               & $2.38$                
                      & $1.80\pm0.13$         & $0.85_{-0.14}^{+0.13}$ & $1.47_{-0.07}^{+0.08}$ \\
$T_{\rm in}$ (keV)      & $ 1.27 $              &  ---                  
                      & $1.38_{-0.06}^{+0.07}$   & $0.36_{-0.04}^{+0.06}$  & $2.00_{-0.05}^{+0.06}$   \\
$R_{\rm in}$ \footnotemark[$\ddagger$]              
                      & $ 51.4 $              &  ---                  
                      & $29.2_{-2.1}^{+2.5}$     & $ \le 88.0$         & $81.9_{-6.5}^{+5.9}$    \\
$L_{\rm disk}$\footnotemark[$\S$]          
                      & $0.62$                &  ---                  
                      & $0.29$                & $\le0.01$           & $1.36$        \\
$p$                   & ---                   &  ---                  
                      & ---                   & ---                 & $0.535_{-0.005}^{+0.004}$ \\
$\Gamma$              & ---                   & $2.17$                
                      & $2.30_{-0.07}^{+0.08}$    & $1.91_{-0.01}^{+0.04}$ &  --- \\
$T_{\rm e}$ (keV)       & ---                   &  ---                  
                      & ---                   & $1.62_{-0.07}^{+0.10}$  &  ---   \\
\hline 
$\chi^2/{\rm dof}$    & $3032.7/1373$         & $2429.7/1373$   
                      & $1535.8/1371$         & $1549.6/1370$ & $1541.8/1372$        \\
\hline 
\multicolumn{6}{@{}l@{}}{\hbox to 0pt{\parbox{145mm}{\footnotesize
\par\noindent
\footnotemark[$*$] Absorption-inclusive 0.5 -- 10 keV flux in $10^{-11}$ ergs cm$^{-2}$ s$^{-1}$
\par\noindent
\footnotemark[$\dagger$] Absorption-corrected 0.5 -- 10 keV luminosity in $10^{39} (\cos i)^{-1}$ ergs s$^{-1}$ 
\par\noindent
\footnotemark[$\ddagger$] 
  The true innermost disk radius in $(\cos i)^{-0.5}$ km.
  The spectral hardening factor of $\kappa = 1.7$
  and the correction factor for the boundary condition of $\xi = 0.412$ 
  are applied in the case of the MCD component,
  while the extended MCD model adopted 
  $\kappa = 3$ \citep{Sim_GRS1915} and $\xi = 0.353$ \citep{Mass_Slimdisk}.
\par\noindent
\footnotemark[$\S$] Absorption-corrected bolometric disk luminosity
in $10^{39} (\cos i)^{-1}$ ergs s$^{-1}$, evaluated in 0.001 -- 100 keV 
\par\noindent
\footnotemark[$\|$] The errors are omitted, because the fit was unacceptable. 
}\hss}}
\end{tabular}
\end{center}
\end{table*}

\subsection{X-ray lightcurve} 
\label{sec:lcurve}
Figure \ref{fig:lcurve} shows
the background-subtracted lightcurves of M33 X-8 
in the soft ($0.5$ -- $2$ keV) and hard ($2$ -- $10$ keV) bands,
where the data from the two front-side illuminated (FI) CCD chips 
(XIS 0 and 3; \cite{XIS}) were summed up.
The source hardness,
plotted in the bottom panel of figure \ref{fig:lcurve},
was simply calculated 
as the ratio of the hard band count rate to the soft band one.

Summed over the two FI CCDs,
we measured the time-averaged signal count rate of M33 X-8 
as $0.77 \pm 0.03$ cts s$^{-1}$ and $0.56 \pm 0.03$ cts s$^{-1}$, 
respectively, in the soft and hard band.
Figure \ref{fig:lcurve} suggests a weak 
but statistically significant intensity variation 
($\chi^2/{\rm dof} = 145.3/37$ and $142.9/37$)
during the Suzaku exposure in the soft and hard ranges 
with a standard deviation of $4.8\%$ and $5.4 \%$, respectively,  
around the average count rate.
The hardness of the object was revealed 
to be relatively unchanged during the Suzaku exposure, 
although the source appears to exhibit a possible spectral softening 
near the end of the exposure.

\subsection{X-ray spectrum} 
\label{sec:spectrum}
\begin{figure*}[t]
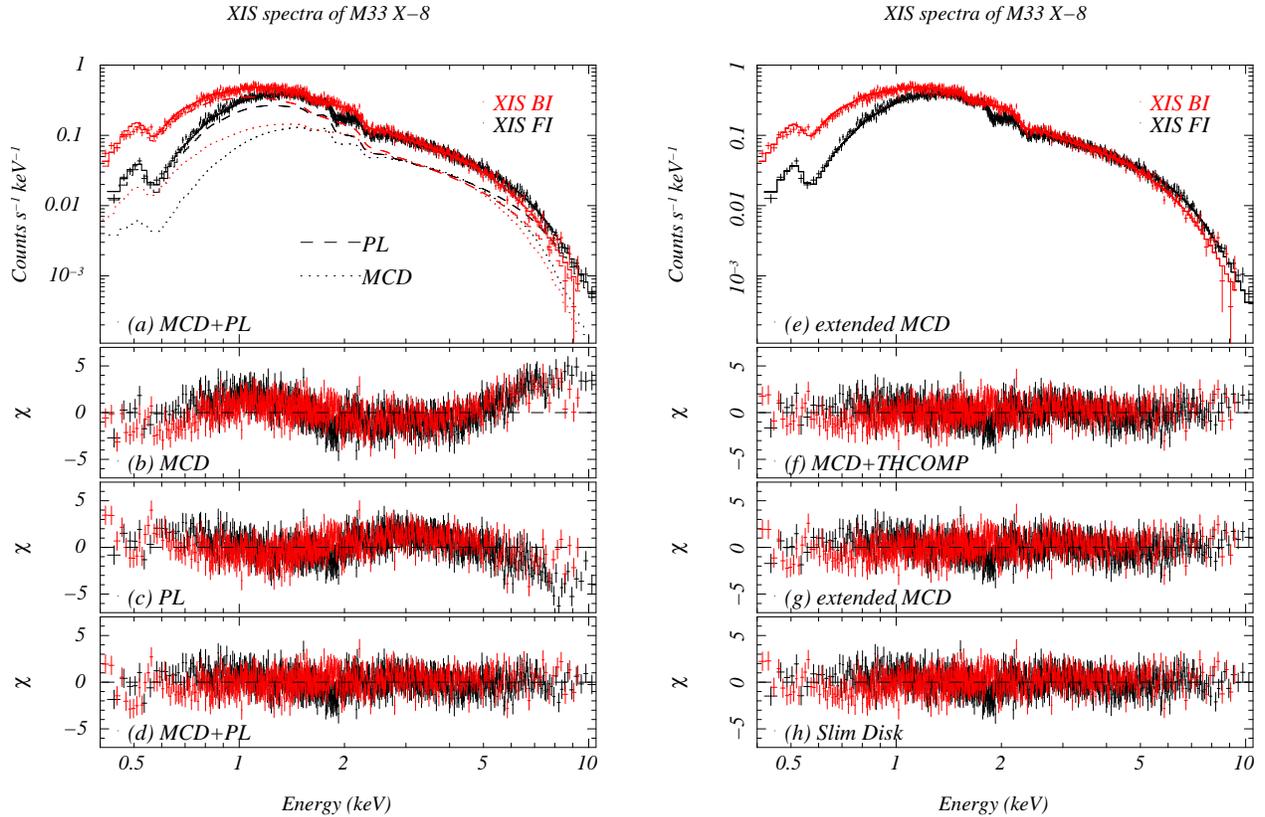

\begin{center}
\FigureFile(80mm,80mm){figure3a.ps}
\hspace{0.5cm}
\FigureFile(80mm,80mm){figure3b.ps}
\end{center}
\caption{Background-subtracted XIS spectrum of M33 X-8,
shown without removing the instrumental response. 
Panels (a) and (e) compares the XIS data 
with the best-fit MCD+PL and extended MCD models, respectively.
The residuals are displayed for 
(b) the best-fit MCD, 
(c) PL, (d) MCD+PL, (f) MCD+THCOMP, (g) extended MCD, 
and (h) slim disk \citep{Slim_Kawaguchi} models, respectively.  
}
\label{fig:spec}
\end{figure*}

Since the change of hardness of M33 X-8
was suggested to be insignificant 
during this observation as shown in figure \ref{fig:lcurve},
we consider only the time-averaged spectrum of the source. 
Figure \ref{fig:spec} shows the background-subtracted XIS spectrum of M33 X-8 
in the range of $0.4$ -- $10$ keV, derived with the FI and 
back-side illuminated (BI; XIS 1, \cite{XIS}) CCD chips,
without removing the instrumental response.
Both the FI and BI spectra were binned 
into energy intervals each with at least 100 events.
In comparison to the previous observations 
with ASCA \citep{M33X8_ASCA}, BeppoSAX \citep{M33X8_SAX}, 
Chandra \citep{M33X8_Chandra} and XMM-Newton (e.g., \cite{M33X8_XMM}),
the signal statistics of the XIS spectrum,
with a total FI and BI signal counts 
of $1.21\times10^{5}$ and $7.47\times10^{4}$ respectively,
seem to be considerably higher. 
The source was found to exhibit a featureless X-ray spectrum,
without any clear sign of emission or absorption structure,
except for instrumental and interstellar ones.

For the purpose of quantifying the spectral properties of M33 X-8, 
we calculated the response matrix functions 
for the individual CCD chips with the HEADAS tool {\tt xisrmfgen},
while the auxiliary response files were generated 
by the tool {\tt xissimarfgen} \citep{xissimarfgen},
assuming a point source at the sky position of M33 X-8.
The {\tt xissimarfgen} tool precisely takes into account 
the reduction of the effective area 
due to contaminants on the optical blocking filters of the XIS,
by referring to the appropriate version of the CALDB (see \S \ref{sec:obs}).
We performed the spectral fitting using the version 12.6.0q
of XSPEC.

As a first step, 
we tried to reproduce the observed XIS spectrum from M33 X-8
with an MCD model or a simple PL one.
A Galactic absorption was adopted with a free hydrogen column density.  
These models are widely used as a standard tool
for analyzing the X-ray spectra of Galactic black hole binaries and ULXs.
As summarized in table \ref{table:spec},
both models were found to be unacceptable
($\chi^2/{\rm dof} = 3032.7/1373$ and $2429.7/1373$
for the MCD and PL models, respectively).
The residual spectra with the MCD and PL models,
shown in panels (b) and (c) of figure \ref{fig:spec}
respectively,
suggest that the observed X-ray spectrum of the source seems 
less convex than the MCD model,
while it is more convex than the simple PL model.

As a more sophisticated description for the X-ray spectrum 
from an accretion disk, the BHSPEC model \citep{bhspec} 
\footnote{This model is available 
in the form compatible with XSPEC from the following web 
{\tt http://heasarc.gsfc.nasa.gov/docs/xanadu/xspec/models/bhspec.html}}
was fitted to the XIS data.
The model assumes 
a fully relativistic accretion disk around a spinning black hole, 
and self-consistently considers 
the vertical structure and radiative transfer within the disk.
The disk spectrum was simulated 
for two representative values 
of the viscosity parameter, $\alpha = 0.1$ and $0.01$.
There are following four free parameters in the model;
the black hole mass $M$, disk luminosity $L_{\rm disk}$,
disk inclination angle $i$
(the face-on configuration corresponds to $i = 0$), 
and dimensionless spin parameter $a$.
Table \ref{table:bhspec} tabulates the resultant parameters.  
Although the fit was slightly improved by the BHSPEC model 
for $\alpha = 0.1$ ($\chi^2/{\rm dof} = 2118.6/1371$)  
in comparison with the MCD or PL model, 
it was still inadequate for interpreting the XIS spectrum of M33 X-8.

\begin{table*}[t]
\caption{Results from the BHSPEC or BHSPEC+THCOMP fitting}
\label{table:bhspec}
\begin{center}
\begin{tabular}{llll}
\hline\hline 
Model           & \multicolumn{2}{l}{BHSPEC \footnotemark[$*$]} 
                    & BHSPEC+THCOMP \footnotemark[$\S$]\\ 
\hline 
$N_{\rm H}$ ($10^{21}$ cm$^{-2}$ )     & $0.41$ & $0.25 $  
                                   & $1.26_{-0.19}^{+0.10}$\\
$M$       ($M_\odot$)               & $8.51$  & $13.0$  
                                   & $25.7_{-1.4}^{+2.0}$\\ 
$\cos i$                           & $> 0.99$ & $0.67$ 
                                   & $0.23_{-0.03}^{+0.07}$\\ 
$a$ \footnotemark[$\dagger$]       & $> 0.79$ & $0.78$ 
                                   & $ a \ge 0.77$\\ 
$L_{\rm disk}$ ($L_{\rm E}$)           & $0.80$  & $0.65$  
                                   & $0.40_{-0.01}^{+0.14}$\\
$\alpha$ \footnotemark[$\ddagger$]  & $0.1$  & $0.01$  
                                    & $0.1$ \\
$\Gamma$                            & ---     & --- 
                                    & $2.21_{-0.11}^{+0.02}$ \\
$T_{\rm seed}$ (keV)                   & ---     & --- 
                                    & $0.21_{-0.04}^{+0.05}$ \\
$T_{\rm e}$ (keV)                     & ---     & --- 
                                    & $\ge 4.5$\\
\hline 
$\chi^2/{\rm dof}$                 & $2118.6/1371 $    & $2448.0/1371$
                                   & $1505.0 / 1367 $ \\
\hline 
\multicolumn{3}{@{}l@{}}{\hbox to 0pt{\parbox{75mm}{\footnotesize
\par\noindent
\footnotemark[$*$] All the errors are omitted.
\par\noindent
\footnotemark[$\dagger$] Dimensionless spin parameter.  
\par\noindent
\footnotemark[$\ddagger$] Fixed parameter in the model.  
\par\noindent
\footnotemark[$\S$] The best-fit parameters are shown only for $\alpha = 0.1$. 
}\hss}}
\end{tabular}
\end{center}
\end{table*}

Next, we examined a combination of an MCD component and a PL one (MCD+PL),
both of which are subjected to a common free absorption.
As shown in panels (a) and (d),
a reasonable fit was derived by the model ($\chi^2/{\rm dof} = 1535.8/1371$)
with the spectral parameters listed in table \ref{table:spec}.
We obtained the absorption-inclusive observed X-ray flux 
as $1.57 \times 10^{-11}$ ergs s$^{-1}$ cm$^{-2}$ in $0.5$ -- $10$ keV. 
The absorption column density,
$N_{\rm H} = (1.80\pm0.13) \times 10^{21}$ cm$^{-2}$,
became higher by a factor of 1.7 than the Galactic value 
($N_{\rm H} = 1.1 \times 10^{21}$ cm$^{-2}$; \cite{NH}).
By extrapolating the best-fit MCD component to the 0.001 -- 100 keV range,
we estimated the absorption-corrected bolometric disk luminosity
as $L_{\rm disk} = 2.9 \times 10^{38} (\cos i)^{-1} $ ergs s$^{-1}$
The innermost disk temperature of the MCD component 
was determined as $T_{\rm in} = 1.38_{-0.06}^{+0.07}$ keV.
The apparent innermost disk radius is evaluated as 
$r_{\rm in} = 24.6_{-1.8}^{+2.0} (\cos i)^{-0.5}$ km, 
based on the equation $L_{\rm disk} = 4 \pi \sigma r_{\rm in}^2 T_{\rm in}^4$ 
\citep{MCD1,MCD2}
where $\sigma$ is the Stefan-Boltzmann constant. 
By taking into account 
a factor $\xi$ to correct the inner disk boundary condition 
and a spectral hardening factor $\kappa$ 
representing the ratio of the color temperature to the effective one,
$r_{\rm in}$ is converted to the true innermost disk radius 
as $R_{\rm in} = \xi \kappa^2 r_{\rm in}$ \citep{ULX_asca}.  
According to the standard manner for the MCD model, 
we adopted $\xi = 0.412$ \citep{xi} and $\kappa = 1.7$ \citep{kappa}.
Thus, the true inner radius was evaluated as 
$R_{\rm in} = 29.2_{-2.1}^{+2.5} (\cos i)^{-0.5}$ km.
We measured the photon index of the PL component 
as $\Gamma = 2.30_{-0.07}^{+0.08}$.
The best-fit model indicates a dominance of the PL component 
over the MCD one,
except for a narrow range of $3$ -- $5$ keV.  

The PL component from the Galactic black hole binaries 
is thought to arise via Comptonization of the disk photons 
in the hot corona around the black hole and accretion disk. 
This motivated us to re-analyze the XIS spectrum 
by replacing the PL component with a thermal Comptonization continuum
described by the {\tt nthcomp} model in XSPEC 
(\cite{thcomp}; hereafter THCOMP).  
The seed photons for the THCOMP component were assumed 
to be supplied by the MCD one. 
As shown in panel (f) of figure \ref{fig:spec},
this MCD+THCOMP model became acceptable ($\chi^2/{\rm dof} = 1549.6/1370$),
yielding the best-fit parameters tabulated in table \ref{table:spec}.
The model resulted in a full dominance of the THCOMP component
in the XIS energy range,
and the upper limit on 
the absorption-inclusive $0.5$ -- $10$ keV flux of the MCD component  
was evaluated as $F_{\rm X} = 1.4 \times 10^{-13}$ ergs cm$^{-2}$ s$^{-1}$ 
which corresponds to only $\sim 1$ \% of the total X-ray flux.
The disk temperature of the MCD component 
(i.e., the seed photon temperature)
was found to be significantly lower than that in the MCD+PL model 
as $T_{\rm in} = 0.36_{-0.04}^{+0.06}$  keV. 
We derived the asymptotic photon index and coronal electron temperature 
of the THCOMP component 
as $\Gamma = 1.91_{-0.01}^{+0.04}$ and 
$T_{\rm e} = 1.62_{-0.07}^{+0.10}$ keV, respectively. 
On the basis of the equation 
\begin{eqnarray*}
\tau = 
\sqrt{ 2.25 + 
\frac{3} {(T_{\rm e}/511~{\rm keV}) [(\Gamma + 0.5)^2-2.25]} }
- 1.5
\end{eqnarray*}
\citep{Comptonization},
the combination of these two quantities
indicates an optically thick corona 
with an optical depth of $\tau \sim 15$.

We also considered a Comptonizing corona surrounding 
an accretion disk around a spinning black hole
by a sum of the BHSPEC and THCOMP components (BHSPEC+THCOMP). 
Because the BHSPEC model is 
not written in terms of the disk temperature, 
it is unable to constrain the seed photon temperature,
$T_{\rm seed}$, of the THCOMP component 
to be consistent with the peak of the BHSPEC spectrum,
through the BHSPEC+THCOMP fitting procedure alone. 
Actually,
although a statistically reasonable fit was derived by the 
BHSPEC+THCOMP model with $T_{\rm seed}$ left free 
(see table \ref{table:bhspec}),
this model got settled to a physically unacceptable situation 
where the spectral peak of the BHSPEC component ($\ge 4.5$ keV) 
highly surpasses the seed photon temperature 
($T_{\rm seed} = 0.21_{-0.04}^{+0.05}$ keV).
In addition, we failed to reconcile the seed photon temperature 
with the BHSPEC peak energy,
in spite of a survey on the seed photon temperature 
in the range of $T_{\rm seed} = 0.3$ -- $2$ keV. 
Therefore, in the present paper,
we have given up to interpret the X-ray spectrum of M33 X-8 
by Comptonization of X-ray photons 
from an accretion disk accompanied with a spinning black hole.

Finally, 
in order to investigate the possibility of the slim accretion disk, 
which is regarded to be important at a high mass-accretion rate,
we adopted a simple extension of the MCD model, 
where the radial temperature distribution on the disk is described 
as $r^{-p}$ with the index $p$ being a positive free parameter 
\citep{pfree_disk}.
Although the model has been widely referred to as a $p$-free disk model,
we regard this nomenclature as somewhat confusing.
We, then, call it as the extended MCD model in the present paper.
The model with $p = 0.75$ is identical to the simple MCD model,
in which a standard accretion disk \citep{standard_disk} is adopted, 
while that with a smaller value of $0.5 < p < 0.75$, 
is considered to approximate phenomenologically 
the X-ray spectra from a slim disk (e.g., \cite{slimdisk}).

Panels (e) and (g) of figure \ref{fig:spec} clearly visualizes 
that the extended MCD model reproduced the observed XIS spectrum of M33 X-8 
well with a goodness of fit, $\chi^2/{\rm dof} = 1541.8/1372$,
similar to that of the MCD+PL or MCD+THCOMP models.
A rather flat temperature profile with $p = 0.535_{-0.005}^{+0.004}$
was indicated by the model.
We estimated the bolometric disk luminosity 
as $L_{\rm disk} = 1.36 \times 10^{39} (\cos i)^{-1} $ ergs s$^{-1}$. 
The innermost disk temperature, $T_{\rm in} = 2.00_{-0.05}^{+0.06}$ keV,
was found to be higher than that in the MCD+PL model.
In the similar manner as for the MCD+PL model, 
we obtained the apparent innermost disk radius 
as $r_{\rm in} = 25.8_{-2.1}^{+1.8} (\cos i)^{-0.5}$ km. 
In the case of the slim disk,
the spectral hardening factor is suggested to become larger 
than that in the standard accretion disk ($\kappa = 1.7$; see above),
due to various effects 
including enhanced electron scattering and Comptonization 
(e.g., \cite{Sim_GRS1915,Slim_Kawaguchi}). 
Here, we adopt $\kappa = 3$ as a typical value for the slim disk,
which was required to reconcile 
the observed high disk temperature of the Galactic microquasar GRS $1915$+$105$ 
with the prediction from the numerical simulation 
on the disk at a high accretion rate \citep{Sim_GRS1915}. 
In addition, the boundary condition is corrected 
by assuming $\xi =0.353$, which was derived by taking into account 
the transonic flow in the pseudo-Newtonian potential \citep{Mass_Slimdisk}.
These yielded the "true" disk radius 
as $R_{\rm in} = 81.9_{-6.5}^{+5.9} (\cos i)^{-0.5} (\xi/0.353) (\kappa/3)^2$ km.

\section{Discussion} 
\label{sec:discussion}
\subsection{Summary of the Suzaku observation} 
In the Suzaku observation conducted on 2011 January 11 -- 13
with an effective exposure of $91.0$ ks,
we successfully acquired the X-ray spectrum of the nearest ULX, M33 X-8, 
with the highest signal statistics in the 0.4 -- 10 keV range. 
During the observation,
the object exhibited no significant change in its spectral hardness, 
with only a weak intensity variation with a standard deviation of $\sim 5$\%.
Without being corrected for the absorption, 
the time-averaged 0.5 -- 10 keV flux of the source was 
measured as $1.57\times 10^{-11}$ ergs s$^{-1}$ cm$^{-2}$ 
(evaluated from the best-fit MCD+PL model).
Re-evaluated from the previous results 
\citep{M33X8_ASCA,M33X8_SAX,M33X8_Chandra,M33X8_XMM2,M33X8_XMM,M33X8_LowTeCompton},  
its 0.5 -- 10 keV flux was found to stay within a relatively narrow range 
of $(1.4$ -- $1.9) \times 10^{-11}$ ergs s$^{-1}$ cm$^{-2}$ over nearly 16 years
\footnote{Here, we neglected several XMM-Newton observations
(ObsID = 0102642001, 0141980501, and 0141980601), 
for which we found inconsistency of more than 20 \% 
in the flux determination between the separate results 
\citep{M33X8_XMM2,M33X8_XMM,M33X8_LowTeCompton}. 
Even if these results are taken into account, 
the observed 0.5 - 10 keV flux varied 
by only a factor of $\sim 2$ in the range of 
$(1.1$ -- $2.4) \times 10^{-11}$ ergs s$^{-1}$ cm$^{-2}$.}. 
Thus, the source X-ray flux during the Suzaku exposure 
is located in the middle of its distribution in the previous observations. 

\begin{figure}[t]
\begin{center}
\includegraphics[angle=-90,width=8cm]{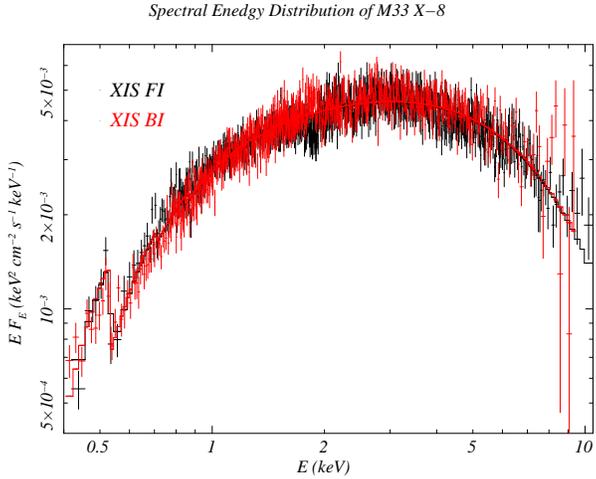}
\end{center}
\caption{
Absorption-uncorrected spectral energy distribution of M33 X-8. 
The XIS data were de-convolved with the best-fit extended MCD model.}
\label{fig:sed}
\end{figure}

Figure \ref{fig:spec} illustrates that 
the time-averaged XIS spectrum of the source was described 
by the MCD+PL, MCD+THCOMP, or extended MCD one,
with the best-fit parameters summarized in table \ref{table:spec},
while the single PL, MCD or BHSPEC model was found to be unacceptable. 
Assuming the best-fit extended MCD model,
the XIS spectrum was de-convolved into the spectral energy distribution 
in the $\nu F_{\rm \nu}$ form, as shown in figure \ref{fig:sed}. 
These results are utilized to infer the nature of M33 X-8 
in the following. 

\begin{table}[t]
\caption{Physical parameters evaluated by the Slim disk model \citep{Slim_Kawaguchi}}
\label{table:slim_disk}
\begin{center}
\begin{tabular}{ll}
\hline\hline 
Parameter                          & Value         \\
\hline 
$N_{\rm H}$ ($10^{21}$ cm$^{-2}$ )     & $1.23_{-0.05}^{+0.06}$ \\
$M$       ($M_\odot$)                & $13.1_{-0.7}^{+0.8}$   \\
$\dot{M}$ ($L_{\rm E}/c^2$)           & $10.5_{-0.7}^{+0.5}$   \\
$\alpha$                            & $ 0.21 \pm 0.05 $      \\
$L_{\rm X}$ (ergs s$^{-1}$)  \footnotemark[$*$]        
                                    & $0.68 \times 10^{39}$       \\
$L_{\rm disk}$ (ergs s$^{-1}$) \footnotemark[$\dagger$]         
                                    & $0.94 \times 10^{39}$ \\
\hline 
$\chi^2/{\rm dof}$                   & $1516.7/1372$\\
\hline 
\multicolumn{2}{@{}l@{}}{\hbox to 0pt{\parbox{65mm}{\footnotesize
\par\noindent
\footnotemark[] The model assumes a face-on disk configuration ($i = 0$). 
\par\noindent
\footnotemark[$*$] Absorption-corrected 0.5 -- 10 keV luminosity. 
\par\noindent
\footnotemark[$\dagger$] Absorption-corrected bolometric disk luminosity evaluated in 0.001 -- 100 keV. 
}\hss}}
\end{tabular}
\end{center}
\end{table}

\subsection{Comptonization scenario} 
\label{sec:Comptonization}
When we adopt the MCD+PL model, 
both the disk temperature and the PL photon index of M33 X-8,
$T_{\rm in} = 1.38_{-0.06}^{+0.07}$ keV and $\Gamma = 2.30_{-0.07}^{+0.08} $,
are found to fall within the range derived in the previous results 
(e.g., \cite{M33X8_ASCA,M33X8_SAX,M33X8_Chandra,M33X8_XMM}). 
Especially, the steep photon index appears to 
be close to those of Galactic black hole binaries in the very high state 
($\Gamma \gtrsim 2.4$; e.g., \cite{GX339_VHS,XTEJ1550_VHS}). 
Actually, 
the very-high-state interpretation was proposed 
for the source by \citet{M33X8_XMM2},
based on the XMM-Newton spectra obtained in a few occasions. 
However, 
we regard that this interpretation encounters several difficulties. 

The first obstacle for the very-high-state scenario 
comes from the high disk temperature of M33 X-8, 
which significantly exceeds that of Galactic black-hole binaries 
in the strongly Comptonized very high state
($T_{\rm in } \ll 1 $ keV), termed intermediate hard state. 
In the case of the Galactic black-hole binaries, 
such a low disk temperature in the very high state is 
explained by a possible disk truncation 
at a radius significantly larger than the last stable circular orbit
\citep{XTEJ1550_VHS}. 
Secondly, the PL dominance 
at energies below the disk temperature $T_{\rm in}$
seems quite unphysical.
The PL component in the very high state is naturally predicted to 
exhibit a sharp low-energy cut-off below the seed photon energy,
because it is basically attributed to 
Comptonization of disk photons within the hot corona,
surrounding the black hole and accretion disk. 

We tried to reconcile these inconsistencies,
by replacing the PL component with the Comptonization continuum 
described by the THCOMP mode,
where the MCD emission was assumed to provide the seed photon source.
The MCD+THCOMP model has eventually arrived at a different solution that 
the disk photons are Comptonized by a low-temperature 
($T_{\rm e} = 1.62_{-0.07}^{+0.10}$ keV) optically-thick ($\tau \sim 15$) material,
which is clearly deviating from the very high state
($T_{\rm e} \gtrsim 20$ keV and $\tau \lesssim 2$; \cite{XTEJ1550_VHS}). 
Recent numerical simulations anticipate that 
extremely high radiation pressure 
at a highly super-Eddington mass accretion rate ($\dot{M} \gg L_{\rm E}/c^2$)
induces a strong outflow from the disk \citep{disk_sim_3D},
which serves a cool ($T_{\rm e} \ll 10$ keV) 
optically thick ($\tau\gg1$) Comptonizing corona \citep{LowTeCorona}.
Such a spectral features are gradually suggested from several 
high-luminosity ULXs (e.g., \cite{UltraluminousState,HoII-X1}). 
In addition, 
\citet{M33X8_LowTeCompton} applied a similar idea 
to the spectral variation of M33 X-8.
However, the high-quality XIS spectrum from the source 
requires the absorption column density for the MCD+THCOMP model 
($N_{\rm H} = 0.85_{-0.14}^{+0.13} \times 10^{21}$ cm$^{-2}$)
to be lower than the Galactic value
($N_{\rm H} = 1.1 \times 10^{21}$ cm$^{-2}$; \cite{NH}). 
We regard such a situation invalid,
considering the absorption within the M33 galaxy and intrinsic to the source.
Therefore, we have concluded that the Comptonization 
is not a realistic scenario for M33 X-8.

\subsection{Slim-disk interpretation} 
\label{sec:slim}
From observational (e.g., \cite{XTEJ1550}) 
and theoretical (e.g., \cite{slimdisk})
studies of Galactic black hole binaries and their accretion disk
in the decade,
the extended MCD model is recognized as a good approximation 
to the X-ray spectrum from a slim accretion disk,
which is thought to be realized at a high X-ray luminosity.
Actually, the model was successfully utilized to interpret 
the X-ray spectra of Galactic black hole binaries, 
at the highest end of its X-ray luminosity or 
in the sub-/trans-Eddington regime, as the slim disk state 
(e.g., XTE J1550-564; \cite{XTEJ1550}).
These were subsequently followed by successful applications 
to the X-ray spectral properties of several ULXs 
(e.g., \cite{M81X9,NGC1313_Suzaku,NGC2403Src3}).

We demonstrated that the XIS spectrum of M33 X-8 was successfully described 
by the extended MCD model. 
The innermost disk temperature and radial temperature gradient,  
$T_{\rm in} = 2.00_{-0.05}^{+0.06}$ keV and $p = 0.535_{-0.005}^{+0.004}$ respectively,  
were both found to agree fairly with those previously inferred from the object 
($T_{\rm in} = 1.4$ -- $2.5$ keV and $p = 0.50$ -- $0.56$; \cite{M33X8_XMM}).  
Especially, the flat radial temperature distribution within the disk 
agrees with the prediction from the slim-disk theory \citep{slimdisk}. 
Therefore, the slim-disk picture seems convincing for the source.

Different from Galactic black hole binaries in the classical high/soft state,
which have been observationally recognized to harbor a standard accretion disk
extending down to the innermost stable circular orbit at $3 R_{\rm S}$,
a mass estimation from the assumption of $R_{\rm in } = 3 R_{\rm S}$ dose not hold
for a black hole with a slim disk, 
where $R_{\rm S}$ is the Schwarzschild radius. 
This is because theoretical simulations on the slim disk widely predict  
that emission from the accretion flow inside the innermost stable circular orbit
becomes not negligible \citep{slimdisk}.
In relation, apparent dependence of the innermost disk radius 
(and hence the luminosity) upon the disk temperature 
is suggested from the numerical simulation \citep{slimdisk} 
as $R_{\rm in} \sim T_{\rm in}^{-1}$ 
(corresponding to $L_{\rm disk} \sim T_{\rm in}^2$). 
Similar properties were actually observed from several ULXs 
within the framework of the standard MCD+PL spectral modeling 
(e.g., \cite{ULX_asca2}).

In order to cope with this problem,
\citet{Mass_Slimdisk} proposed a simple method.
They fitted the numerically simulated X-ray spectra of a slim disk 
to the extended MCD model. 
They found from some representative values of black hole mass, 
$M = (10$ -- $100) M_\odot$, 
and accretion rate, $\dot{M} = (1$ -- $1000) L_{\rm E}/c^2$,
that a correction factor needs to be applied 
to convert the “apparent” mass $M_{\rm X}$ derived 
from the hypothesis of $R_{\rm in } = 3 R_{\rm S}$ 
to the true one as $M/M_{\rm X} = 1.2$ -- $1.6$.
Here, we follow the technique to estimate the black hole mass of M33 X-8,
assuming that the object accompanies a slim accretion disk.

The innermost disk radius of M33 X-8, 
$R_{\rm in} = 81.9_{-6.5}^{+5.9} (\cos i)^{-0.5}$ km
(here and hereafter dependence on $\kappa$ and $\xi$ is neglected for clarity),
measured from the XIS spectrum with the extended MCD fitting
gives the apparent black hole mass as 
$M_{\rm X} = (9.2\pm0.7) (\cos i)^{-0.5} M_\odot$. 
Adopting the conversion factor of $M/M_{\rm X} \sim 1.2$
(for $M = 10 M_\odot$; \cite{Mass_Slimdisk}), 
the true mass of M33 X-8 is estimated as $M \sim 11 (\cos i)^{-0.5} M_\odot $.
Since the mass corresponds to the Eddington luminosity of 
$L_{\rm E} \sim 1.7 \times 10^{39}  (\cos i)^{-0.5}$ ergs s$^{-1}$, 
the source is indicated to shine at the sub-Eddington luminosity of 
$L_{\rm X}/L_{\rm E} \sim 0.4 (\cos i)^{-0.5}$
or
$L_{\rm disk}/L_{\rm E} \sim 0.8 (\cos i)^{-0.5} $.
Considering the observational fact that 
the Galactic black hole binaries are known to exhibit 
the slim-disk-like X-ray properties typically 
at the Eddington ratio of $L_{\rm X}/L_{\rm E} = 0.3$ -- $1$ 
(e.g., \cite{XTEJ1550,GROJ1655_VHS}),
the slim-disk interpretation is thought to become self-consistent for M33 X-8.  

A number of authors (e.g., \cite{M82X1_slimdisk,ULX_slim,M51_ULX}) 
utilized the X-ray model spectrum from a slim accretion disk 
generated from the numerical simulation by \citet{Slim_Kawaguchi}
\footnote{This model is available 
in the form compatible with XSPEC from the following web 
{\tt http://heasarc.gsfc.nasa.gov/docs/xanadu/xspec/models/slimdisk.html}.}
to claim a super-Eddington interpretation for several ULXs, 
including M33 X-8 \citep{M33X8_slimdisk}. 
These have driven us to investigate the XIS spectrum with this model. 
The model takes into account 
the effects of electron scattering and general relativity 
in detail (with the parameter Model ID set at 7). 
We have to note 
that only the face-on disk geometry ($i = 0$) is supposed in the model. 
By leaving free $M$, $\dot{M}$ and the viscosity parameter $\alpha$,
the model yielded a reasonable fit ($\chi^2/{\rm dof} = 1516.7/1372$)
as shown in the panel (h) of figure \ref{fig:spec}. 
The physical quantities derived 
from the slim disk model are tabulated in table \ref{table:slim_disk}.
The black-hole mass, $M = 13.1_{-0.7}^{+0.8} M_\odot$,
was found to nearly agree with the result from the extended MCD model. 
The model requires the viscosity parameter of $\alpha = 0.21 \pm 0.05 $, 
which is not deviating far from 
that usually adopted for ULXs ($\alpha = 0.1$; e.g., \cite{ULX_slim,M51_ULX}).
\citet{M33X8_slimdisk} reported to prefer a lower mass of $ M < 10 M_\odot$, 
by applying the same model to the XMM-Newton spectra of M33 X-8
obtained in the three occasions.
However, we regard that their result is possibly an artifact 
due to a rather low value of the disk viscosity ($\alpha < 0.01$)
adopted in their analysis. 
It is important to note that the luminosity of the ULX 
is indicated to be significantly less than the Eddington limit 
as $L_{\rm X}/L_{\rm E} \sim 0.3$ or $L_{\rm disk}/L_{\rm E} \sim 0.5$,
even though the source is suggested to be accreting 
at a super Eddington rate as $\dot{M} = 10.5_{-0.7}^{+0.5} L_{\rm E}/c^2$.

We regard these results as basically supportive of the slim disk idea 
with the sub-Eddington luminosity
for M33 X-8. 
However, we have to be careful about a few 
theoretical and observational uncertainties.
The black hole mass derived from the extended MCD model  
is subject to the ambiguity in the spectral hardening factor $\kappa$. 
The numerical model on the slim accretion disk \citep{Slim_Kawaguchi} 
is recommended to be calibrated 
by the Galactic black hole binaries with a known mass,
although they are known to exhibit rarely the slim disk properties. 
In addition, it is expected from recent numerical studies 
(e.g., \cite{disk_sim_3D,LowTeCorona}) that 
the slim disk tends to accompany a strong outflow. 
The X-ray spectrum from the slim disk is possible to be contaminated 
from Comptonized X-ray photons produced in the outflow. 

\section{Implication to the nature of the ULXs}
By making most of the Suzaku XIS spectrum 
with the highest signal quality ever recorded, 
we propose the slim disk scenario for the nearest ULX, M33 X-8. 
In addition, both the empirical extended MCD model 
and the complicated numerical simulation on the slim accretion disk 
have guided us to the conclusion that 
the source hosts a black hole with a mass of 
$M\gtrsim 10 M_\odot$ (for $i = 0$). 
If the disk viewing angle is taken into consideration, 
the mass will become larger as $ \propto (\cos i)^{-0.5}$ 
in the simplest manner neglecting the effects of the general relativity.
One of the most remarkable points from this investigation is that 
the source is found to be radiating at the sub-Eddington luminosity,
in spite of its super-Eddington mass accretion rate. 
A similar observational support was revealed by \citet{NGC2403Src3},
to the sub- or trans-Eddington luminosity from the slim accretion disk 
by analyzing the Suzaku data for the ULX, 
Source 3 in the nearby galaxy NGC 2403. 

A sub/trans-Eddington luminosity is 
not surprising from the theoretical point of view, 
even in the case of the super critical accretion flow 
($\dot{M} \gg L_{\rm E} / c^2$),
since the luminosity tends to saturate 
due to strong advection and photon trapping. 
Actually, a disk luminosity of $L_{\rm disk}/L_{\rm E} \sim 3$ at most
was anticipated from the numerical simulation on the slim disk  
at the significantly super critical accretion rate 
as $\dot{M} = 1000 L_{\rm E} / c^2 $ (e.g., \cite{Mass_Slimdisk}). 
In summary, we have asserted that 
it is not necessary to invoke the highly super-Eddington luminosity, 
in order to understand the ULX behavior. 
By extrapolating this consideration 
to the ULXs more luminous than M33 X-8, 
it is natural to conclude that 
there are intermediate mass black holes with a mass of $M \gtrsim 100 M_\odot$ 
among those with $L_{\rm disk} \gg 10^{40}$ ergs s$^{-1}$. 

The bandpass of most operating X-ray instruments limited below 10 keV 
frequently prevented us to disentangle clearly 
the degeneracy between the candidate spectral models
(e.g., the slim disk or cool Comptonization model). 
Actually, the capability of the hard X-ray spectrum was certified 
by the recent Suzaku HXD detection of the hard X-rays up to $\sim20$ keV
from the ULX, M82 X-1 \citep{M82X1_Suzaku}, 
since the result have succeeded 
in specifying the source spectrum as a Comptonized emission 
from a black hole with a mass of $M = (100$ -- $200) M_\odot$. 
Moreover, 
even in the case that a source is confirmed to host a slim accretion disk,
a hard band X-ray spectrum above 10 keV is expected to be useful 
to eliminate the possible uncertainty due to the contamination 
from the strong outflow (see \S \ref{sec:slim}).   
Therefore, we expect that 
a wide-band X-ray spectrum in the range of 0.3 -- 80 keV 
to be derived with the next-generation Japanese X-ray observatory 
ASTRO-H \citep{ASTRO-H} will be inevitably 
play a crucial role to settle the nature of the ULXs. 

We thank the anonymous referee for her/his supportive guidance 
to improve remarkably the present paper.
We are grateful to all the members of the Suzaku team,
for the successful operation and calibration.
This research is partially supported by the
MEXT Grant-in-Aid for Young Scientists (B) 22740120 (N. I.).


\end{document}